\documentclass[aps,twocolumn,showpacs]{revtex4-1}
\usepackage{graphicx}
\usepackage[all]{xy}
\usepackage{amsmath}
\usepackage{amssymb}
\newcommand{\be}{\begin{equation}}
\newcommand{\ee}{\end{equation}}
\newcommand{\ben}{\begin{eqnarray}}
\newcommand{\een}{\end{eqnarray}}
\newcommand{\bes}{\begin{subequations}}
\newcommand{\ees}{\end{subequations}}

\newcommand{\bb}{\bibitem}
\begin{document}
\title{Construction of new scalar field models from the standard $\phi^4$ theory}

\author{D. Bazeia,$^{a,b}$ M.A. Gonz\'alez Le\'on,$^c$ L. Losano,$^{a,b}$, J. Mateos Guilarte$^d$, and J.R.L. Santos,$^a$}
\affiliation{{\small $^a$Departamento de F\'{\i}sica, Universidade Federal da Para\'{\i}ba, Jo\~ao Pessoa, PB, Brazil\\$^b$Departamento de F\'\i sica, Universidade Federal de Campina Grande, Campina Grande, PB, Brazil\\
$^c$Departamento de Matem\'atica Aplicada and IUFFyM, Universidad de Salamanca, Spain\\
$^d$Departamento de F\'{\i}sica Fundamental and IUFFyM, Universidad de Salamanca, Spain}}
\email[Corresponding Autor \\ {\it Email Address:\ }]{guilarte@usal.es}

\date{\today}
\begin{abstract}
In this work we study models described by a single real scalar field in two-dimensional space-time, using the deformation procedure to propose and investigate new families of models and their kink solutions.
\end{abstract}


\maketitle

\section{Introduction}

Our goal is the construction of relativistic scalar field models and the analysis of their extended solutions. We use the deformation procedure, which was put forward in Ref.~\cite{blm}-\cite{blm1}, and profusely applied in \cite{dd+} in a diversity of contexts. The deformation procedure
relies on a function, called the deformation function, which responds for the construction of the new, deformed model. Here, however, we choose the deformation function as a composed function of the scalar field under investigation. This novelty nicely brings new interesting families of models, together with their corresponding extended solutions.

The composed deformation function greatly enlarges the capabilities of the method allowing also the deformation of singular solutions of the starting model into regular solutions of the deformed model. Another interesting new result in this work is that all the potentials obtained through the composed deformation can be written in a factorized form, immediately allowing the identification of the absolute minima of the new potential terms.

The deformation method introduced in \cite{blm} connects two distinct models of real scalar fields in two-dimensional space-time, characterized respectively by the Lagrangians
\bes \ben
{\cal L}=\frac12\partial_\mu\chi\partial^\mu\chi-U(\chi)\label{1a}
\\
{\cal L}_d=\frac12\partial_\mu\phi\partial^\mu\phi-V(\phi)\label{1b}
\een\ees
where $\chi$ and $\phi$ are the scalar fields and
$U(\chi)$ and $V(\phi)$ are the potential terms, which specify each one
of the two models. The key ingredient
 is an invertible function $f=f(\phi),$ the deformation function, from which we link the
model of (\ref{1a}) with the \lq\lq deformed" model of (\ref{1b}) by
relating the two potentials $U(\chi)$ and $V(\phi)$ in the very
specific form \be V(\phi)=\frac{U(\chi\to f(\phi))}{(df/d\phi)^2}
\ee
This allows showing that if the starting model has static
solution $\chi(x)$ which obeys the first-order equations
${d\chi}/{dx}=\pm\sqrt{2U(\chi)} \label{1c}$,
and the equation of motion ${d^2\chi}/{dx^2}={dU}/{d\chi} $,  then, the
deformed model has static solution given by
$\phi(x)=f^{-1}(\chi(x)),$ which obeys ${d\phi}/{dx}=\pm\sqrt{2V(\phi)} \label{1d}$,
and ${d^2\phi}/{dx^2}={dV}/{d\phi} $.
The proof was already given in Ref.~{\cite{blm}}.

The effectiveness of the deformation method
 in the search for topological and non-topological
defects in (1+1)D scalar field theory demands, first, that the static solutions of eq.(\ref{1c}) are  known.
The second subtle point is an shrewd choice of deformation function such that
$V(\phi)$ has a finite set of degenerated minima:
${\phi^i} (i=1, \cdots,N)$, and the analytical solution $\phi_S(x)=f^{-1}[\chi_S(x)]$
of  eq.(\ref{1d}) complies with: $\lim_{x\to
-\infty}\phi_S(x)=\phi^i$ and $ \lim_{x\to\infty}\phi_S(x)=\phi^j \,
, \, \, j=i+1, i, i-1 \quad $ If $j=i\pm 1$ we talk of
topological defects named kinks/anti-kinks, and if $j=i$ we find, besides the
classical minima, non-topological defects called lumps.

\section{Composed deformation functions}
\label{sec:3}

In Ref.~\cite{blm1}, some of us applied the deformation method starting from the
standard $\chi^4$ model and choosing
\be
\chi=f(\phi)=\cos(a\arccos\phi-M\pi) \, ,
\ee
as deformation functions, where $a$ and $M$ are
integer or half-integer numbers.  In this
paper we extend the family of deformation functions by allowing
composed functions $f[g(\phi)]$ of the form
\be
\label{df}\chi=f[g(\phi)]=\cos(a\arccos[g(\phi)]-M\pi)\, .
\ee
Appropriate choices of the function $g(\phi)$ will provide us with
new scalar field models having analytically solvable first-order
differential field equations, which support new topological
(kink-shape, double kink-shape) and/or non-topological (bell-shape, sugar loaf-shape)
defect structures as classical static solutions of the corresponding
field equations.

The starting model is described by the potential \be
U(\chi)=\frac12\;(1-\chi^2)^2 \ee where we are using dimensionless
field and coordinates. We fix the center of the defect at the origin
($x_0=0$) to get, for $|\chi|\leq 1$, the finite energy regular kink
solution $ \chi_1(x)=\pm\tanh(x) $ and, for $|\chi|\geq 1$, the
infinite energy singular kink solution $ \chi_2(x)=\pm\coth(x) $.
Like in Ref.~ \cite{blm1}, the parameter $M$ leads to two
distinct families of models: for $M$ integer, the deformed potential
can be written in the form \be\label{vs1}
{V}_{\sin}^a(\phi)=\frac{1}{2a^2}\frac{(1-g^2(\phi))}{(g'(\phi))^2}\sin^2\left(a\;\arccos(g(\phi))\right)
\ee where $g'(\phi)=dg/d\phi$. However, for $M$ half-integer we get
\be\label{vc1}
{V}_{\cos}^a(\phi)=\frac{1}{2a^2}\frac{(1-g^2(\phi))}{(g'(\phi))^2}\cos^2\left(a\;\arccos(g(\phi))\right)
\ee Using half-integer and integer values
for the parameter $a$ the number of vacua of the new model is fixed at will.

By this procedure we identify families of potentials which present
static solutions of the general form \be\label{sol1}
{\phi}_S(x)=g^{-1}\left(\cos\left[{(\eta(x)+
M\;\pi})/{a}\right]\right)\ee where $g^{-1}$ is the inverse function
of $g(\phi)$, and $\eta(x)$ is either $\theta_1(x)$ or $\theta_2(x)$,  both $\in [0,\pi]$ given by
\ben
\theta_1(x)&=&\arccos(\tanh(x))\,, \label{teta1}\\
\theta_2(x)&=&\arccos(\coth(x))\,.\label{teta2}
\een

Firstly, we start with \eqref{vs1} and $g(\phi)=\phi$. Here the potentials are, for $a$ odd,
\be\label{vsino}
{V}_{\sin}^a(\phi)=\frac{1}{2a^2}\;\prod_{j=1}^{(a+1)/2} \left(1-\frac{\phi^2}{{Z_j^{a}}^2}\right)^2\,,
\ee
and for $a$ even,
\be\label{vsine}
{V}_{\sin}^a(\phi)=\frac{1}{2}\;\phi^2\;\prod_{j=1}^{a/2} \left(1-\frac{\phi^2}{{Z_j^{a}}^2}\right)^2\,,
\ee
where $Z_j^{a}=\cos[(j-1)\pi/a]$. Also, for $a$ half-integer we get
\be\label{vsins}
{V}_{\sin}^a(\phi)=\frac{1}{4a^2}(1-\phi)(1-\phi^2)\;\prod_{j=1}^{a-1/2} \left(1+\frac{\phi}{{\tilde{Z}_j^{a}}}\right)^2\,,
\ee
where $\tilde{Z}_j^{a}=\cos[(2j-1)\pi/2a]$.
Now, if we start with \eqref{vc1} and $g(\phi)=\phi$, the potentials are, for $a$ odd,
\be\label{vcoso}
{V}_{\cos}^a(\phi)=\frac{1}{2}\;\phi^2(1-\phi^2)\prod_{j=1}^{(a-1)/2} \left(1-\frac{\phi^2}{ {{\tilde{Z}}_j^{a 2}}}\right)^2\,,
\ee
and, for $a$ even,
\be\label{vcose}
{V}_{\cos}^a(\phi)=\frac{1}{2a^2}\;(1-\phi^2)\;\prod_{j=1}^{a/2} \left(1-\frac{\phi^2} { {\tilde{Z}}_j^{a 2}}\right)^2\, ,
\ee
whereas, for $a$ half-integer we have ${V}_{\cos}^a(\phi)={V}_{\sin}^a(-\phi)$.

Note that, the choice $g(\phi)=\phi$ gives the polynomial potentials already investigated in Ref.~\cite{blm1}, but here the formula for $a$ half-integer, Eq.~\eqref{vsins}, is factorized.

\section{Families of models for $g(\phi)=\phi^r$}
\label{sec:4}

Consider the case  $g(\phi)=\phi^r$, where $r={n}/{m}$
is a positive rational number, the ratio of two nonzero natural
numbers, i.e., $n,m\in{\mathbb N}^*$. From \eqref{sol1}, the static solutions of the $V^{a}_{\sin}(\phi)$ and
$V^{a}_{\cos }(\phi)$ are given by
\be\label{uroot}
{\phi}_S^k(x)=s(r)\cos^{1/r}\left[({\eta(x)+M\;\pi})/{a}\right]
\quad ,
\ee
where $M=k-1$ for the $sine$ family, or $M=(2k-1)/2$ for the $cosine$ family, and $k$ is a positive natural number, but only
a few values of $k$ produce different solutions depending on $a$. The symbol $s(r)$ in (\ref{uroot}) is
defined as follows: If $r={2p}/{(2q-1)}$, $p,q\in{\mathbb N}^*$, $s(r)$ amounts to take the $\pm$ sign and the
modulus before extracting the odd root in Eq.\eqref{uroot}:
\be
{\phi}_S^k(x)=\pm\left|\cos^{2q-1}\left[(\eta(x)+M\;\pi)/{a}\right]\,\right|^\frac{1}{2p}
\quad . \label{uroot1} \ee For any other value of $r$ the symbol is simply the unity: $s(r)=1$.
This is because there are two real even roots and only one real odd root.
Because $r$ is positive, the deformation
function $\chi=g(\phi)=\phi^r$ maps the range $|\chi|\leq 1$ in the
range $|\phi|\leq 1$, and all the zeros of the deformed potential
are between $-1\leq\phi\leq 1$. Then, to obtain finite energy static
solutions of the deformed model we need to start from the regular
kink solution $\eta(x)=\theta_1(x)$, as given by Eq.~\eqref{teta1}.

We note that a general characteristic of the models generated by the deformation function \eqref{df} with
$g(\phi)=\phi^r$, is that the number of topological and non-topological static solutions is determinate
only by $a$, as described below. The potentials can be written in polynomial form, for $a$ integer or half-integer, as we show below.

\subsection{The $sine$ family of models for $a$ integer}

Here the family of models $V^{a,r}_{\sin}$ is investigated
for the specific case of $a$ being an integer. The
polynomial form of ${V}^{a,1}_{\sin}$, with its zeros (and
multiplicities) is known, and so performing the deformation
$g(\phi)=\phi^r$ we have, for $a$ odd
\be {V}_{\sin}^{a,r}(\phi)=\frac{1}{2a^2r^2}\;\phi^{2-2r} \,
\prod_{j=1}^{(a+1)/2}
\left(1-\frac{\phi^{2r}}{{Z_j^{a}}^{2}}\right)^2\,,\ee
and for $a$ even
\be {V}_{\sin}^{a,r}(\phi)=\frac{1}{2r^2}\;\phi^2\,
\prod_{j=1}^{a/2} \left(1-\frac{\phi^{2r}}{{Z_j^{a}}^{2}}\right)^2\,,
\ee
where $Z_j^a=\cos\left[(j-1)\pi/a\right]$ under the restriction that the potential be real and
nonsingular. Henceforth, if $a$ is odd we can take only $r\leq 1$.
The $sine$ potential $V^{a,r}_{\sin}(\phi)$ can be written in terms
of the Chebyshev polynomials of second kind in the $\phi^r$ variable:
\bes \ben
V^{a,r}_{\sin}(\phi)&=&\frac1{2a^2r^2}\;\phi^{2-2r}(1-\phi^{2r})^2\;U^2_{a-1}(\phi^r)\,,
\\
\nonumber
\\
U_a(\theta)&=&{\sin[(a+1)\arccos\theta]}/{\sin[\arccos\theta]}\label{cheb2}\,.
\een \ees
The explicit forms of $V^{a,r}_{\sin}(\phi)$, for
${a=1,2,3}$ are given by
\be\label{vsin1n}
V^{1,r}_{\sin}(\phi)=\frac1{2r^2}\;\phi^{2-2r}\;(1-\phi^{2r})^2\,,
\ee
\be V^{2,r}_{\sin}(\phi)=\frac1{2r^2}\;\phi^2\;(1-\phi^{2r})^2\,,
\ee
\be
V^{3,r}_{\sin}(\phi)=\frac{8}{9r^2}\;\phi^{2-2r}\;(1-\phi^{2r})^2\left(\frac14-\phi^{2r}\right)^2\,,
\ee
which illustrate this new family of models.
We see that, in the cases for $r>1$ integer  and
$r>1/2$ half-integer, we have to take $a$ even.

The defects analytically described by the formula (\ref{uroot}) are classified in
three distinct types: topological kink, non-topological bell-shape lump, and topological double kink. The defect classes
depend on the potentials $V^{a,r}_{\sin}(\phi)$.
We have noticed in  Ref.~\cite{blm1}, that in the case for $r=1$ there are two classes
of models: for $a$ odd they are $\phi^4-$like potentials -- no zero
at the origin -- and for $a$ even they are $\phi^6$-like models -- having a zero at the origin.
Hereafter,  the new potentials are described and compared with their predecessors  in Ref.~\cite{blm1} case by case.

\begin{figure}[ht]
\includegraphics[{height=02.2cm,width=08cm,angle=00}]{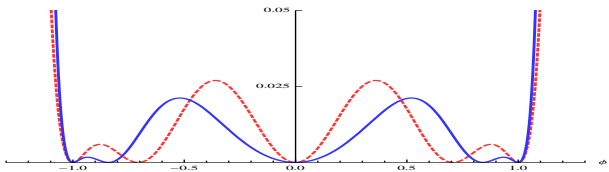}
\caption{Plots of $V^{4,1}_{\sin}(\phi)$ and $V^{4,2}_{\sin}(\phi),$ depicted with dashed (red) and solid (blue) lines, respectively.}\label{FIG1}
\end{figure}

For $a=$ even, $r=n$ or $r=n/m$, $n$ integer, and $m$ odd, the potentials are non-negative and symmetric with respect to $\phi=0$, like the
$\phi^6$- model, have a zero at the origin, see Figures \ref{FIG1} and \ref{FIG4}.
The vacua and the static solutions are
\be
\phi_v^j=\pm\left|\cos\left[(j-1)\pi/a\right]\,\right|^\frac{1}{r} \, , \, \, \, j=1, \, \dots \, , 1+{a}/{2}\,,
\ee
\be \phi_S^k(x)=\pm\Big|\cos\Big[\arccos(\tanh(x))+(k-1)\pi)/{a}\Big]\,\Big|^\frac{1}{r},\ee
$k=1, \, \dots \, , a$.
There are $a+1$ vacua and $a$ pairs of kink/anti-kink, both for $n$ even and odd. All the defects
are topological kink/anti-kink, interpolating between consecutive vacua of the potential.

\begin{figure}[ht]
\includegraphics[{height=02.2cm,width=08cm,angle=00}]{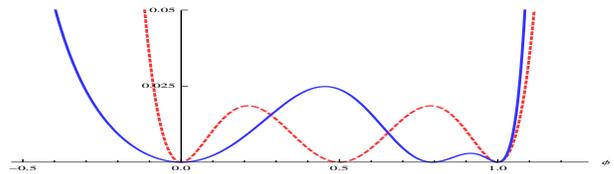}
\caption{Plots of $V^{4,1/2}_{\sin}(\phi)$ and $V^{4,3/2}_{\sin}(\phi),$ depicted with dashed (red) and solid (blue) lines, respectively.}\label{FIG2}
\end{figure}

\begin{figure}[ht]
\includegraphics[{height=02.2cm,width=08cm,angle=00}]{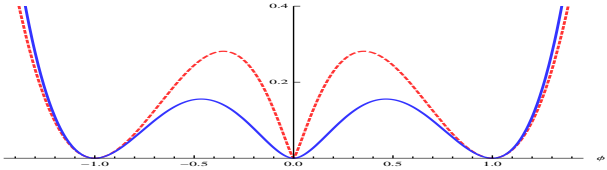}
\caption{Plots of $V^{1,1/3}_{\sin}(\phi)$ and $V^{2,1/3}_{\sin}(\phi),$ depicted with dashed (red) and solid (blue) lines, respectively}\label{FIG4}
\end{figure}

For $a$ even and $r=n/2>1/2$ half-integer, the potentials are also non-negative but non symmetric with respect to the zero at the origin,
that is a minimum of $V$. These potentials are not $\phi\to -\phi$ invariants and all their critical points are non-negative, see Figure \ref{FIG2}.
The vacua and the static solutions are
\be \phi_v^j=\left(\cos^{2}\left[(j-1)\pi/a\right]\right)^{\frac{1}{n}}\, , \,\,
j=1,2, \dots , 1+ {a}/{2}\,,\ee
\be \phi_S^k(x)=\left(\cos^{2}\left[({\arccos(\tanh(x))+(k-1)\pi})/{a}\right]\right)^{\frac{1}{n}}\, \, \, , \ee $k=1,2, \dots , a$. There are thus $a/2+1$ non-negative vacua and $a/2$ couples of topological kink/anti-kink defects.

For  $a$ odd and $r={1}/{2}$,  the potentials of the $sine$ family
are non-negative only for $\phi\geq0$, and the zero at $\phi=0$ is
not a critical point.

The vacua and the static solutions are respectively
\be
\phi_v^j=\cos^2\left[({j-1})\pi/a\right] \, , \, \, j=1,2, \dots , ({a+1})/{2}\,,
\ee
\be
\phi_S^k(x)=\cos^2\left[({\arccos(\tanh(x))+(k-1)\pi})/{a}\right]\, \, ,
\ee
$k=1,2, \dots, a$. There are $(a+1)/2$ vacua and a total of $a$ defects - one non-topological lump and $(a-1)/2$ couples of topological kink/anti-kink. The solutions corresponding to
$k<(a+1)/2$ are topological kinks interpolating between two consecutive local minima.
For $k>(a+1)/2$ we find topological anti-kinks that connect the vacua in the opposite sense. The remaining solution $k=(a+1)/2$ asymptotically behaving as $\phi=\cos^2\left[(a-1)\pi/2a\right]$ both at $x=\pm\infty$, is a non-topological lump: in
the mechanical analogy, the associated NTK trajectory in the
potential $V^{a,1/2}(\phi)$ starts at $x=-\infty$ from the
\lq\lq maximum" $\phi=\cos^2\left[(a-1)\pi/2a\right]$, bounces back at the \lq\lq turning" point
$\phi=0$ and \lq\lq finally" arrives at $\phi=\cos^2\left[(a+1)\pi/2a\right]=\phi=\cos^2\left[(a-1)\pi/2a\right]$ at $x=\infty$.

\begin{figure}[ht]
\includegraphics[{height=02.2cm,width=08cm,angle=00}]{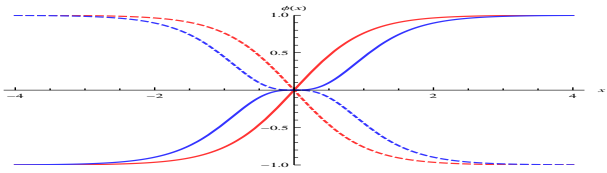}
\caption{Plots of kink (solid line) and anti-kink (dashed line) solutions of $V^{1,1/3}_{\sin}$ (blue) and $V^{1,1}_{\sin}$ (red).
The solutions of $V^{1,1/3}_{\sin}$ have the form of double kink/anti-kink.}\label{FIG5}
\end{figure}

Finally, for $a$ odd and $r=n/m$ non integer, $m$ odd, and $n=1,2, \dots ,m-1 $, the potentials are
non-negative and symmetric with respect to the origin, where $V=0$, see Figure \ref{FIG4}.
A very interesting novelty arise in these cases: the origin is not a vacuum of $V$,
$[{dV}/{d\phi}]_{\phi=0}$
does not exist for $r>1/2$, and $[{d^2V}/{d\phi^2}]_{\phi=0}\rightarrow\infty$ for $r<1/2$. This generate a defect connecting the two closest neighbor minima near the origin, named double kink-like solution. The vacua and the static solutions are
\be
\phi_v^j=\pm\left|\cos^{m}\left[(j-1)\pi/a\right]\,\right|^\frac{1}{n} \, , \, \, \, j=1, \, \dots \, , ({a+1})/{2}\,,
\ee
\be \phi_S^k(x)=\pm\left|\cos^{m}\left[({\arccos(\tanh(x))+(k-1)\pi})/{a}\right]\,\right|^\frac{1}{n},\ee
where $k=1, \, \dots \, , a$, and  $k\neq({a+1})/{2}$ for kink/anti-kink defect, and
\be
\phi_S^k(x)=\pm sg(x) \left|\cos^{m}\left[(\arccos(\tanh(x))+(k-1)\pi)/a\right)\right|^\frac{1}{n}
\ee
where  $sg(x)=x/|x|$, and $k=(a+1)/2$ for double kink/anti-kink defect.
There are $a+1$ vacua and $a$ couples of topological defects - (a-1) couples of kinks/anti-kinks, and one couple of double kink/anti-kink around the origin - both for $n$ even and odd. For instance, let us take
$n=1$, the $k=(a+1)/2$ and $k=a+2$ kinks
skip the origin and connect the closest minima to $\phi=0$: $\phi=\cos^{m}\left[({a-1})/{2a}\right]$
and $\phi=\cos^{m}\left[({a+1})/{2a}\right]$. The character is thus topological joining two different minima but they have the shape of a double kink. This special form of defect was introduced in Ref.~\cite{bmm}, and one nicely see its appearance again in the current context. In Figure \ref{FIG5} we plot the double kink and double anti-kink
solutions of the sine potential for $a=1$ and $r=1/3$.

\subsection{The $cosine$ family of models for $a$ integer}

Here is investigate the family of models given by $V^{a,r}_{\cos}$
for the specific case of $a$ being an integer. As before, the polynomial form
of ${V}^{a,1}_{\cos}$, with its zeros (and multiplicities), is also known.
Performing the deformation $g(\phi)=\phi^r$ we get to the new potentials:
for $a$ odd,
\be {V}_{\cos}^{a,r}(\phi)=\frac{1}{2r^2}\;\phi^2 (1-\phi^{2r}) \,
\prod_{j=1}^{\frac{a-1}{2}} \left(
1-\frac{\phi^{2r}}{{Z_j^{a}}^2}\right)^2\,, \ee
and for $a$ even,
\be {V}_{\cos}^{a,r}(\phi)=\frac{1}{2a^2r^2}\;\phi^{2-2r}
(1-\phi^{2r}) \, \prod_{j=1}^{\frac{a}{2}} \left(
1-\frac{\phi^{2r}}{{Z_j^{a}}^2}\right)^2\,,
\ee
where $Z_j^a= \cos\left(\frac{2j-1}{2a}\pi\right)$. Again the
restriction that the potential be real and nonsingular means that
for $a$ even we can take only $r\leq 1$. The $cosine$ potentials
$V^{a,r}_{\cos}(\phi)$ are given by Chebyshev polynomials of the first
kind in the $\phi^r$ variable: \bes \ben
V^{a,r}_{\cos}(\phi)&=&\frac1{2a^2r^2}\;\phi^{2-2r}(1-\phi^{2r})\;T^2_a(\phi^r)\,,\\
T_a(\theta)&=&\cos[a\arccos\theta]\label{cheb1}\,. \een \ees

The explicit forms of the potentials $V^{a,r}_{\cos}(\phi)$ for
${a=1,2,3}$ are given by
\be
V^{1,r}_{\cos}(\phi)=\frac1{2r^2}\;\phi^2\;(1-\phi^{2r})\,,
\ee
\be
V^{2,r}_{\cos}(\phi)=\frac1{2r^2}\;\phi^{2-2r}(1-\phi^{2r})\left(\frac12-\phi^{2r}\right)^2\,,
\ee
\be
V^{3,r}_{\cos}(\phi)=\frac8{9r^2}\;\phi^2\;(1-\phi^{2r})\;\left(\frac34-\phi^{2r}\right)^2\,,
\ee
which illustrate this new family of models.
We see that, in the cases for $r>1$ integer and $r>1/2$
half-integer, we can take only $a$ odd.

Here, the defects analytically described by the formula (\ref{uroot}) can be classified in
three types: topological kink, non-topological bell-shape lump, and topological double kink. The defect classes
depend on the potentials $V^{a,r}_{\cos}(\phi)$. We recall that in the case for $r=1$,
there are two classes of models: for $a$ odd they are inverted $\phi^4-$like models
-- having a zero at the origin --, and for $a$ even they are inverted $\phi^6-$like models.
Hereafter, the new potentials are described and compared with their predecessors in the third work in Ref.~\cite{blm1} case by case.

For $a=$ odd, $r=n$ or $r={n}/{m}$, $n$ integer, and $m$ odd, the potentials are non-negative only for $|\phi|\leq 1$,  and
the zeros $\phi=\pm1$ are not critical points, like the
inverted $\phi^4-$ models -- having a zero at the origin.

The vacua and the static solutions are
\be
\phi_v^j=\pm\left|\cos\left[({2j-1})\pi/2a\right]\,\right|^\frac{1}{r} \, , \, \, \, j=1, \, \dots \, , ({a+1})/{2}\,,
\ee
\be
\phi_S^k(x)=\pm\left|\cos\left[({2\,\arccos(\tanh(x))+(2k-1)\pi})/{2a}\right]\,\right|^\frac{1}{r},
\ee
where $k=1, \, \dots \, , a-1$ for kink/anti-kink defect, and $k=a$ for lump defect.
There are $a$ vacua and $2a$ defects - $(a-1)$ couples of topological kink/anti-kink and two non-topological lumps, both for $n$ even and odd.

For $a$ odd and $r=n/2>1/2$ half-integer, the potentials are non-symmetric and
non-negative only for $\phi\leq1$, and the zero at $\phi=1$ is
not a critical point, all their critical points are non-negative.
The vacua and the static solutions are given by
\be \phi_v^j=\left(\cos^{2}\left[({2j-1})\pi/2a\right]\right)^{\frac{1}{n}}\, , \,\,j=1,2, \dots , ({a+1})/{2}\,,\ee
\be \phi_S^k(x)=\left(\cos^{2}\left[({2\,\arccos(\tanh(x))+(2k-1)\pi})/{2a}\right]\right)^{\frac{1}{n}}\, \, \, , \ee where $k=1,2, \dots , a-1$, for kink/anti-kink defect, and $k=a$, for lump defect.
There are $(a+1)/2$ non-negative vacua and $a$ defects - $(a-1)/2$ couples of topological kink/anti-kink and one non-topological lump.

For $a$ even and $r={1}/{2}$,  the potentials of the $cosine$ family
are non-negative only for $0\leq\phi\leq1$, and the zeros at $\phi=0,1$ are
not critical points, all their critical points are non-negative.
The vacua and the static solutions are respectively
\be
\phi_v^j=\cos^2\left[({2j-1})\pi/2a\right] \, , \, \, j=1,2, \dots , {a}/{2}\,,
\ee
\be
\phi_S^k(x)=\cos^2\left[(2\,\arccos(\tanh(x))+(2\,k-1)\pi)/2a\right]\, \, ,
\ee
where $k=1,2, \dots, (a-1)$ $\forall\,k\neq\,a/2$, for kink/anti-kink defect, and $k=a/2\,\,\text{or}\,\,a$, for lump defect. There are $a/2$ vacua and a total of $a$ defects - two non-topological lumps and $(a-2)/2$ couples of topological kink/anti-kink.

Finally, for $a$ even and $r=n/m$ non integer, $m$ odd, and $n=1,2, \dots ,m-1 $, the potentials are
non-negative only for $|\phi|<1$ and symmetric with respect to the origin, where $V=0$,  and
the zeros $\phi=\pm1$ are not critical points.
In this case, the origin is not a vacuum of $V$,
$[{dV}/{d\phi}]_{\phi=0}$ does not exist for $r>1/2$, and $[{d^2V}/{d\phi^2}]_{\phi=0}\rightarrow\infty$ for $r<1/2$. Again, this generates defects connecting the two closest minima to the origin, in the form of double kink  and double anti-kink. The vacua and the static solutions are
\be
\phi_v^j=\pm\left|\cos^{m}\left[({2j-1})\pi/2a\right]\,\right|^\frac{1}{n} \, , \, \, \, j=1, \, \dots \, , {a}/{2}\,,
\ee
\be \phi_S^k(x)=\pm\left|\cos^{m}\left[({2\,\arccos(\tanh(x))+(2k-1)\pi})/{2a}\right]\,\right|^\frac{1}{n},\ee
where $k=1,2, \dots, (a-1)$ $\forall\,k\neq\,a/2$, for kink/anti-kink, $k=a$ for lump defects, and
\be \phi_S^k(x)=\pm\frac {x}{|x|}\left|\cos^{m}\left[\frac{2\,\arccos(\tanh(x))+(2k-1)\pi}{2a}\right]\right|^\frac{1}{n}\ee
where $k={a}/{2}$, for double kink/anti-kink defect.
There are $a$ vacua and $2a$ defects -- $(a-2)$ couples of topological kink/anti-kink, and one couple of topological double kink/anti-kink around the origin, and two non-topological lumps -- both for $n$ even and odd.

We can also choose $a$ half-integer. This gives two new families of models, which can be studied as before.
Moreover, other possibilities for $g(\phi)$ can also be chosen, in particular we can consider the case  $g(\phi)=1/\phi^r$, where $r={n}/{m}$. As before, $r$
is a positive rational number, the ratio of two nonzero natural
numbers, i.e., $n,m\in{\mathbb N}^*$. The investigation follows the same steps we just introduced,  so we omit it here.

\section{Superpotentials and stability}
\label{sec:6}

In general, when the potential is non negative, it is possible to introduce superpotentials $W=W(\phi)$ such that
$
V(\phi)=\left({dW}/{d\phi}\right)^2/2\,.
$
This is the case for the $sine$ family of potentials with $a$ integer. However, in the other cases, for the $cosine$ family of potential with $a$ integer, and for the $sine$ and $cosine$ families of potential with $a$ half-integer, the potentials may be non negative. Nevertheless, we can follow the lines of \cite{ablm}  introducing superpotentials, for both topological and non topological solutions, and their energies.

Particularly, for $r=1$, in the case of the $sine$ and $cosine$ families, and $a$ integer or half-integer, the superpotentials can be written in terms of Chebyshev polynomials as
\bes
\ben
W^{a,1}_{sin}(\phi)&=&\frac{1}{a^2(a^2-4)}[(a^2(1-\phi^2)-2)\;T_a(\phi)\nonumber\\
&&-2a\phi(1-\phi^2)\;U_{a-1}(\phi)]\;,
\een
\ben
W^{a,1}_{cos}(\phi)&=&\frac{\sqrt{1-\phi^2}}{a^2(a^2-4)}[(a^2(1-\phi^2)-2)\;U_{a-1}(\phi)\nonumber\\
&&+2a\phi\;T_{a}(\phi)]\;,
\een
\ees
for $a\neq2$, and
\bes
\be
W_{sin}^{2,1}(\phi)\, =(2\phi^2-\phi^4)/4\,,
\ee
\be
W_{cos}^{2,1}(\phi) =\left((3\phi-2\phi^3)\sqrt{1-\phi^2}+\arcsin(\phi)\right)/8\,.
\ee
\ees
In the case of the $sine$ family,  for $r\neq 2$ we have
\ben
W^{1,r}_{sin}(\phi)&=&\frac{\phi^{2-r}}{r} \left(\frac{\phi ^{2r}}{r+2}+\frac{1}{r-2}\right)\,,\\
W^{2,r}_{sin}(\phi)&=&\frac{\phi^2}{4 r }\left(\frac{\phi ^{2 r}}{r+1}-1\right)\,.
\een
For $r=2$, we have
\be
W^{1,2}_{sin}(\phi)=(\phi^4-4\ln|\phi|)/8
\ee
\be
W^{3,2}_{sin}(\phi)=(2\phi^8-5\phi^4+4\ln|\phi|)/24\,.
\ee

In the case of the $cosine$ family, for $r\neq 2$ we find
\ben
W^{1,r}_{cos}(\phi)&=&\frac{\phi^{2}}{2r(2+r)} \Big(2\sqrt{1-\phi^{2r}}+r\nonumber\\
&&\times\,\,_2F_1\left[{1}/{2},{1}/{r},1+1/r,\phi^{2r}\right]\Big)\,,\\
W^{2,r}_{cos}(\phi)&=&\frac{\phi^{2-r}}{8r(1+r)(r-2)} \Big(2\sqrt{1-\phi^{2r}}\nonumber\\&&\times\left(1+r+(r-2)\phi^{2r}\right)+(2r-1)\phi^{r-2}\nonumber\\
&&\times\,\,\,B\left[\phi^{2r},{1}/{2}+{1}/{r},1/2\right]\Big)\,,
\een
where $_2F_1$ is the Gaussian Hypergeometric function and $B$ the Euler incomplete beta function. For $r=2$, however, we have
\ben
W^{2,2}_{cos}(\phi)&=&\frac1{24}\sqrt{1-\phi^4}\,(5-2\phi^4)+\frac18\ln(\phi^2)\nonumber\\
&&-\frac18\ln\left(2+2\sqrt{1-\phi^4}\right)\,.
\een

In general, the superpotential simplify the calculations. Particularly, for computing the energy associated with the corresponding static solution $\phi_k$ we can write, for the kinklike solutions connecting minima $Z_k$ and $Z_{k+1}$:
$
E^{a,r,k}(\phi_k)=|W^{a,r}(Z_{k})-W^{a,r}(Z_{k+1})|\;,
$
and for the lumplike solution around the minimum  $Z_k$:
$
E^{a,r,k}(\phi_k)=2\;|W^{a,r}(Z_{k})-W^{a,r}(\phi(x=0))|\;.
$
\begin{figure}[ht]
\vspace{0.2cm}
\includegraphics[{height=02cm,width=08cm,angle=00}]{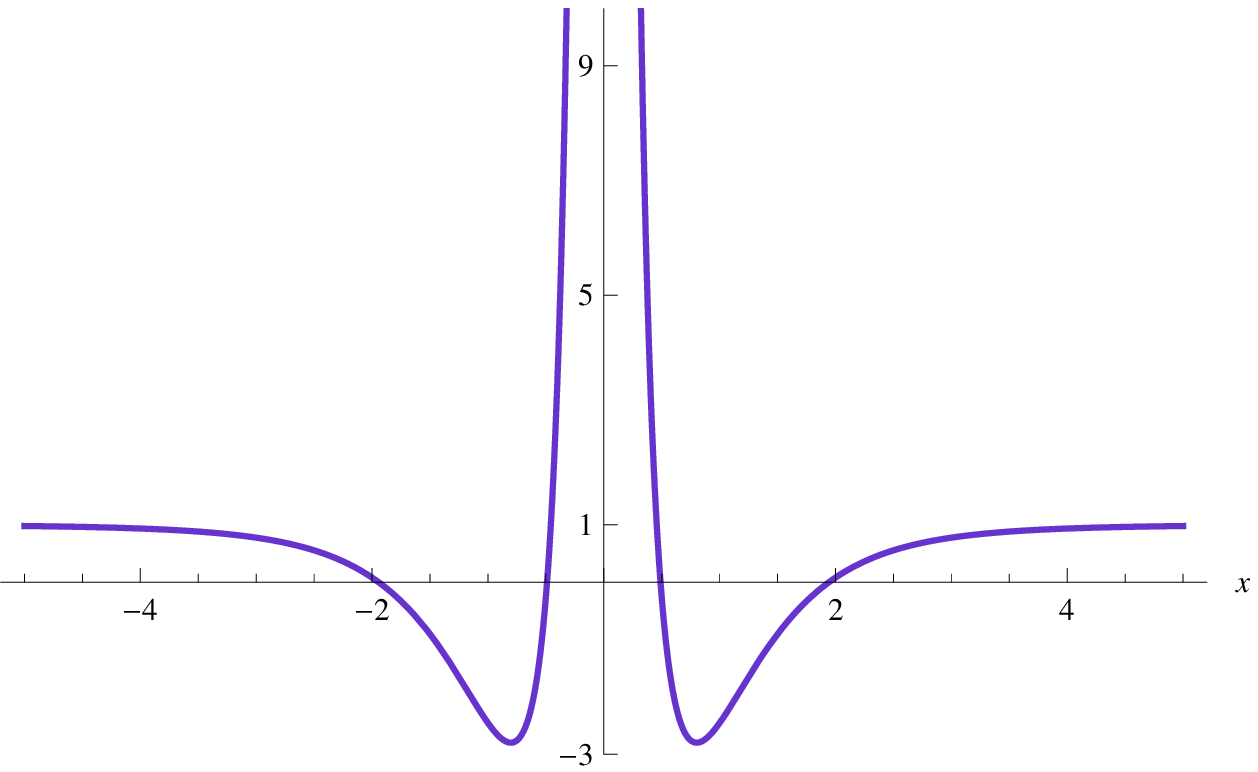}
\includegraphics[{height=02cm,width=8cm,angle=00}]{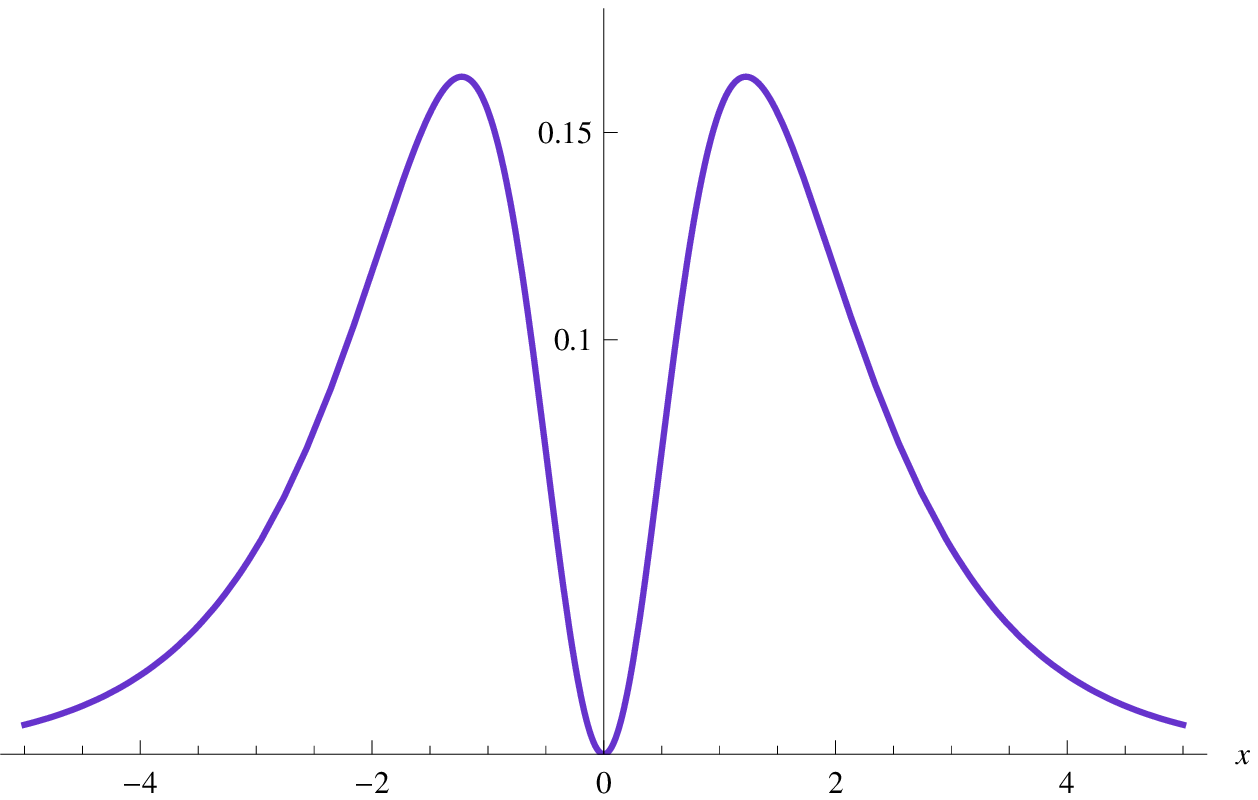}\vspace{0.3cm}
\caption{The potential well $U(x)$ (upper panel) and the translation mode (lower panel), as a
function of $x$}\label{FIG15}
\end{figure}
Regarding the problem of linear stability most of the new solutions
obtained in this work are either ordinary kinks interpolating
between consecutive vacua or classical lumps
connecting a vacuum at $x=-\infty$ with itself at $x=+\infty$. We
have profusely studied the stability of these solutions in previous
papers; ordinary kinks are stable and classical lumps are
unstable, and there is no need to repeat here the same
calculations and arguments. There is a third type of kink that arise
in the new models considered in this work when there are cuspidal
points in the potential. The newly found solutions resemble double
kinks and their stability properties are unclear. Therefore, we
shall only discuss the stability of double kinks choosing as a
completely generic example in this class the potential $V^{2,\frac{1}{3}}_{cos}(\phi)$, and the double kinks:
$
\phi_S(x)=\pm{\sinh^3({x}/{2})}/\cosh^{\frac32}(x)\,.
$
The Schr$\ddot{\rm o}$dinger operator governing the small
fluctuations around these kinks is of the form $K=-{d^2}/{d
x^2}+U(x)$, where the potential well is given by
$
U(x)=1+{1}/{2\sinh^2 (x/2)}-{5}/{2\cosh (x)}-{35}/{4\cosh^2 (x)}\,.
$
Things are more clear in the Figure \eqref{FIG15} where $U(x)$ is depicted. One notices the important limits $\lim_{x\to\pm\infty}
U(x)=1$, $\lim_{x=0}U(x)=+\infty$ and realizes the qualitative similarity with the Lennard-Jones potential \cite{lj} of molecular physics.

The main difference is that the $U(x)$ well looks like the Lennard-Jones well (living only in the positive real half-line) plus its specular image with respect to the ordinate axis, defined as a whole on the full real line. In fact, the ground state wave function of zero energy is the translational mode $\psi_0(x)=d\phi_K / dx = 3 \sinh ({x}/{2})\, \sinh(x) / 4 \cosh^{\frac52}(x) $.

There is no negative energy eigenfunction because the zero of the translational
mode wave function is not strictly a node, see Fig.~\eqref{FIG15}. There is no change of sign in the wave function which is simply telling us that the center of the double kink, where the field reaches the value of zero, cannot be perturbed, since it would cost infinite energy. The other, higher in energy, eigenfunctions are totally reflecting scattering waves with the threshold at $k^2=1$. Either coming from the left or from the right the incoming waves are sent back by the infinite wall at the origin which is perfectly opaque. There are no kink fluctuations that cross the center of the double kink. One spectrum like this is very peculiar but ensures the stability of the double kink: there is no kink fluctuation with negative energy.

{\bf{Acknowledgements}}.
The authors would like to thank CAPES/Nanobiotec and CNPq, Brazil, the FCT project CERN/FP/116358/2010, Portugal, and the Spanish Ministerio de Educaci\'on y Ciencia,  grant FIS2009-10546, for partial financial support.


\end{document}